\documentclass[aps,pre,twocolumn,floatfix]{revtex4-2}

\usepackage{epsf}
\usepackage{subfigure}
\usepackage{amsmath}
\usepackage{amssymb}
\usepackage{bm}
\usepackage{graphicx}
\usepackage{epstopdf}
\newcommand{\be}{\begin{equation}}
\newcommand{\ee}{\end{equation}}
\newcommand{\bea}{\begin{eqnarray}}
\newcommand{\eea}{\end{eqnarray}}

\newcommand{\parent}[1]{\left( #1 \right)}

\begin{document}

\title{Geometric Foundation of Nonequilibrium Transport: A Minkowski Embedding of Markov Dynamics}

\author{David Andrieux}

\noaffiliation

\begin{abstract}
We introduce a unified framework for analyzing Markov dynamics by linking nonequilibrium thermodynamics with information geometry. Using the symmetrized Kullback-Leibler divergence, we reveal an intrinsic Minkowski structure in the parameter space that naturally separates symmetric elements, governed by kinetic activities and forces, from the antisymmetric behavior, characterized by thermodynamic currents and affinities. This formulation offers a systematic approach to probe the transport properties of Markov systems.
\end{abstract}

\maketitle

\vskip 0,25 cm

\section*{Context and objectives}

Many biological, chemical and physical mesoscale systems are well-described by a Markov dynamics \cite{NP77, H05}.
To gain insight into their dynamical and thermodynamical properties, it is crucial to understand their variations across the parameter space.
However, the space of Markov models has no obvious intrinsic structure. 
For example, multiple combinations of parameters can lead to the same affinities and currents, creating ambiguity and arbitrariness in their nonequilibrium response \cite{H05, A24}. 
Analysis of parameter sensitivity also shows that most models are 'sloppy', in the sense that they have a hierarchy of eigenvalues spanning multiple scales \cite{GEA07, MEA13}. 

Information geometry offers a lens to address these challenges. 
In information geometry, the space of parameters (e.g., for probability distributions or Markov dynamics) is analyzed by introducing a divergence as a measure of the separation between models \cite{N05, A16}.
In the linear regime, the parameter space has a natural structure based on the Fisher information metric, which is linked to equilibrium fluctuations \cite{SC12, ZS12}.

Outside the linear regime, however, the structure of the parameter space and its connection to thermodynamics has remained largely unexplored.
One notable exception are dynamical equivalence classes, which provide a 'hidden' structure in the parameter space induced by their transport properties \cite{A12}. These equivalence classes uniquely characterize the nonquilibrium conditions and parameters of the system, providing a well-defined and symmetric nonequilibrium theory \cite{A22, A25}.

In a different context, Teoh et al. demonstrated that the symmetrized Kullback-Leibler divergence plays a special role in vizualizing probabilistic models \cite{QCS19, TQS20}. 
The symmetrized KL divergence induces a Minkowksi structure that preserves the distance between distributions. 
This approach was used to analyze the distributions of states in systems such as ideal gases and 2-dimensional Ising models.

This paper brings these different concepts and results together. 
Following the approach of Teoh et al. \cite{TQS20}, I show that the space of Markov models possesses a Minkoswki structure induced by the symmetrized Kullback-Leibler divergence.
In this Minkowski space, the embedding coordinates disantangle the system's symmetric and antisymmetric components. 
The former is expressed in terms of kinetic activities and forces while the latter coincides with the dynamical equivalence classes and are expressed in terms of thermodynamic currents and affinities.
These findings provide a unified and systematic way to study nonequilibrium Markov systems and their dynamical properties.

\section{The space of Markov chains}

\subsection{Markov chain dynamics}

We consider discrete time Markov chains on a finite state space of size $N$ such that $\sum_j P_{ij} = 1$ (an extension of our results to continuous-time is straightforward).
We assume that the chain is irreducible and that if a transition $i \rightarrow j$ exists then its reverse also exists, i.e. $P_{ij} > 0$ implies $P_{ji} > 0$. Under these conditions, the chain $P$ admits a unique stationary state
\bea
\pi_j = \sum_i \pi_i P_{ij} 
\eea
with $\sum_i \pi_i =1$.

The Markov chain $P$ can be depicted as a graph $G$ with $N$ states, $S$ self-transitions, and $E$ transitions or edges $e \equiv i \longleftrightarrow j$ between states. 
We denote the space of all possible chains with this graph structure by $\Lambda_G$.

\subsection{Thermodynamics of Markov chains}

A system is at thermodynamic equilibrium if the detailed balance conditions $\pi_i P_{ij} = \pi_j P_{ji}$ are satisfied. 
Outside equilibrium, steady-state transport is characterized by the currents
\bea
J_{ij} = \pi_i P_{ij} - \pi_j P_{ji} \, .
\label{Jij}
\eea
The currents $J_{ij}$ must satisfy Kirchoff's law $\sum_{j} J_{ij} = \sum_{i} J_{ij} = 0$ at each node. 
Therefore, any current $J_{ij}$ can be expressed as a linear combination of a subset of independent currents $J_\alpha$ \cite{S76}. 

In tandem with the currents, the thermodynamic forces or affinities are measured by the breaking of detailed balance along cyclic paths $c = (i_1, i_2, \cdots, i_n)$ as
\bea
A_c = \ln \parent{ \frac{P_{i_1i_2}\cdots P_{i_ni_1}}{P_{i_1i_2}\cdots P_{i_ni_1}} } \, .
\eea
The affinities vanish at equilibrium where detailed balance is satisfied. 
Similar to the currents, only a subset $A_\alpha$ of these affinities are independent \cite{S76}. 

Both the currents and affinities are antisymmetric under time-reversal:
\bea
J^*_{ij} = - J_{ij} \quad {\rm and} \quad A^*_{ij} = - A_{ij} \, ,
\eea 
where the time-reversed chain is given by
\bea
P^*_{ij} = P_{ji} \frac{\pi_j}{\pi_i} \, .
\eea 
The time-reversed chain has the same steady state $\pi$ as the original dynamics.
$P$ is at equilibrium if $P^* = P$ .

In turn, the symmetric components of the dynamics are described by the activities \cite{FN03}
\bea
Y_{ij} = \pi_i P_{ij} + \pi_j P_{ji} \, .
\label{Yij}
\eea
The analoguue of the affinities are the "kinetic forces" 
\bea
\bar{X}_{ij} = \ln (P_{ij}P_{ji}) \, .
\eea
Both the activities and kinetic forces are symmetric under time-reversal:
\bea
Y^*_{ij} = Y_{ij} \quad {\rm and} \quad \bar{X}^*_{ij} = \bar{X}_{ij} \, .
\eea

Note that the activities are not all independent since $\sum_{ij} Y_{ij} = \sum_e Y_e =1$, where $e$ denotes the set of undirected transitions $i \longleftrightarrow j$.
One edge $r$ can thus be eliminated as $Y_r = 1 - \sum_{e \neq r} Y_e$, generating effective forces 
\bea
X_e = \bar{X}_e - \bar{X}_r \, .
\eea

\subsection{Representations of Markov chains}

Multiple descriptions of the spaces of Markov chain exist, e.g. using exponential families \cite{N05, A16, WW21} or rotational representations \cite{A82, K06}.
Here a 'thermodynamic decomposition' will provide a useful way to describe the space $\Lambda_G$ \cite{A12}.
In this decomposition, a chain $P$ is determined by its equilibrium component $P^e$ and its affinities $A_\alpha$ as
\bea
P = s \left[ P^e \circ Z(\pmb{A}) \right] \, .
\label{P.decomp}
\eea
The mapping $s$ transforms a positive matrix into a stochastic one as
\bea
s[Q] = \frac{1}{\rho} {\rm diag}(\beta)^{-1} \, Q \, {\rm diag}(\beta) \, ,
\eea
where $\rho$ is the largest eigenvalue of $Q$ and $\beta$ its corresponding right eigenvector. 

The component $P^e$ is given by
\bea
P^e = s[(P \circ P^*)^{(1/2)}] \, ,
\eea 
where $P^{(1/2)}$ denotes the Hadamard or elementwise exponentiation.
$P^e$ satisfies detailed balance, $(P^e)^* = P^e$.  
In turn, the matrix 
\bea
Z_{ij} \equiv \begin{cases} \exp \parent{\pm A_\alpha/2}  & \text{if $i\rightarrow j = \alpha$ in the $\pm$ direction,} \\
                1 & \text{otherwise,}
                \end{cases}\
\nonumber
\eea
captures the antisymmetric (nonequilibrium) components of $P$. 

From this decomposition, the full space of Markov models in $\Lambda_G$ is obtained by varying the equilibrium elements $P^e$ and the affinities $A_\alpha$. 


\section{Embedding in Minkowski space}

The thermodynamic decomposition (\ref{P.decomp}) offers a way to navigate the space $\Lambda_G$. 
However, $\Lambda_G$ still lacks a natural geometric structure or notion of distance.

For exponential families of probability distributions, Teoh et al. showed that the symmetrized Kullback-Leibler divergence $D_S$ offers a natural notion of separation \cite{TQS20}.
Using that divergence \cite{FN01}, an exponential family of probability distributions with $K$ parameters can be embedded isometrically in a $K+K$-dimensional Minkowski space. 
In this Minkowski space, the family of models can be visualized as control parameters are varied while preserving the degree of separation between probability distributions \cite{TQS20}. 

Here we adapt this approach to discrete-time Markov chains. 
The symmetrized Kullback-Leibler divergence between two Markov chains reads
\bea
D_S ( P , P') &=& D ( P | P') +  D ( P' | P) \nonumber \\
&=& \sum_{ij} \left[ \pi_i P_{ij} - \pi'_i P'_{ij} \right] \ln \frac{P_{ij}}{P'_{ij}} \geq 0  \, , \quad
\label{DsKL}
\eea
where
\bea
D ( P | P') =  \sum_{ij} \pi_i P_{ij} \ln \frac{P_{ij}}{P'_{ij}} \geq 0
\eea 
is the Kullback-Leibler divergence and $\pi$ and $\pi'$ are the stationary distributions of $P$ and $P'$, respectively. 

Next, we recast the divergence (\ref{DsKL}) in a more useful form. 
As shown in the Appendix, inserting the thermodynamic decompositions of $P$ and $P'$ into (\ref{DsKL}) leads to
\bea
D_S ( P , P') &=& \frac{1}{2}  \sum_{\alpha} (A_\alpha-A_\alpha') (J_\alpha-J_\alpha') \nonumber \\
&+& \frac{1}{2}\sum_{e} (X_e -  X_e') (Y_e-Y_e')  \, .
\label{DSKLb}
\eea

The expression (\ref{DSKLb}) constitutes our first main result. 
It expresses the symmetrized divergence between two models $P$ and $P'$ in terms of thermodynamic quantities. 
The first contribution captures the antisymmetric aspects of the dynamics and is expressed in terms of the thermodynamic current and affinities, while the second contribution reflects the changes in the symmetric part of the dynamics.

To parametrize the space $\Lambda_G$, we now introduce the coordinates
\bea
S_\alpha^{\pm} = \frac{1}{2} ( A_\alpha \pm J_\alpha ) \quad {\rm and} \quad S_e^{\pm} = \frac{1}{2} ( X_e \pm Y_e ) 
\label{icoord}
\eea
based on the kinetic and thermodynamic variables.
Using these coordinates, the divergence (\ref{DSKLb}) takes the form
\bea
D_S (P,P') &=& \frac{1}{2} \sum_{\alpha} \left[ (\Delta S^+_\alpha)^2 - (\Delta S^-_\alpha)^2 \right]  \nonumber\\
&+& \frac{1}{2} \sum_{e} \left[ (\Delta S^+_e)^2 - (\Delta S^-_e)^2 \right] \geq 0 \, , 
\label{Mink}
\eea 
where $\Delta S^{\pm} = S^{\pm}(P) - S^{\pm}(P')$. 
From expression (\ref{Mink}) we see that the coordinates $S^+$ are spacelike while $S^-$ are timelike.

The embedding (\ref{icoord})-(\ref{Mink}) constitues our second main result.
The space $\Lambda_G$ of model instances forms a $M$-dimensional manifold, called the {\it model manifold}, in an $M+M$ Minkowski space. 
Here $M$ corresponds to the number of independent currents plus the number of edges and self-transitions.
Its maximal value is $2E+S-N$; in practice symmetries can reduce the number of dimensions. This will be illustrated on a molecular motor example in Section \ref{MM}.

Using the coordinates (\ref{icoord}), the set of Markov chains can be vizualized as control parameters are varied, preserving the separation (\ref{Mink}) between models.


\section{Geometry and transport properties of the model manifold}

In this section we explore the geometry of the model manifold.
Contrary to special relativity where spacetime has 1 time dimension and 3 spatial dimensions, here the Minkowski space has $M$ time and $M$ spatial dimensions. 
Notions such as light cones and causal structures thus require a different interpretation. 
Nonetheless, the kinetic and thermodynamic properties of the model instances directly emerge from this underlying geometry. 	 

To study the model manifold, we look at the coordinates $e$ and $\alpha$ separately. 
The two types of coordinates indeed behave differently under time-reversal:
\bea
(S^{\pm}_e)^* = S^{\pm}_e, \quad \, (S^{\pm}_\alpha)^* = - S^{\pm}_\alpha\, .
\eea 
That is, the coordinates $S^{\pm}_e$ are symmetric whereas the $S_\alpha$ are antisymmetric under time-reversal.
This behavior leads to different nonequilibrium transport properties.


\subsection{Light cone of the antisymmetric coordinates $\alpha$}

Let's first consider the light cone $(\Delta S_\alpha^{+})^2 = (\Delta S_\alpha^{-})^2$ for the antisymmetric variables $\alpha$. 
Using coordinates (\ref{icoord}), the light cone of $P$ is given by the dynamics $P'$ satifying
\bea
(A_\alpha - A'_\alpha) (J_\alpha -J'_\alpha) = 0 \, .
\eea
This relation is satisfied if $A'_\alpha=A_\alpha$ or $J'_\alpha=J_\alpha$, i.e., the light cone of variable $\alpha$ has two branches given by the isoaffinities and isocurrent manifolds.

On the joint light cone of all variables $\alpha$, the divergence (\ref{DSKLb}) simplifies to 
\bea
D_S ( P , P') &=& \frac{1}{2} \sum_e \left[ (\Delta S^+_e)^2 - (\Delta S^-_e)^2 \right] \nonumber \\
&=& \frac{1}{2}\sum_{e} (X_e -  X_e') (Y_e-Y_e')  \, . \nonumber
\nonumber
\eea
The separation between two models on this joint light cone is expressed in terms of the systems' activities and kinetic forces only \cite{FN04}.

Note that the coordinates $S^{\pm}_\alpha$ still vary along the isoaffinity or isocurrent manifolds. 
Only on the equilibrium manifold do the coordinates $S^{\pm}_\alpha$ remain constant ($S^{\pm}_\alpha = 0$ in that case). 
The equilibrium manifold is thus special in the sense that it is entirely described by the coordinates $S^{\pm}_e$.

In terms of transport, varying parameters on the light cone changes a system's fluctuations and dissipation while keeping its output (set of currents) or driving forces (affinities) constant.

\subsection{Light cone of the symmetric coordinates $e$}

We now consider the light cone $(\Delta S_e^{+})^2 = (\Delta S_e^{-})^2$ for the symmetric variables $e$. 
The light cone is now given by the relation 
\bea
(X_e - X'_e) (Y_e -Y'_e) = 0 \, .
\eea
Therefore, the two branches of the light cone are given by the isokinetic manifold $X_e = X'_e$ and the isoactivity manifold $Y_e = Y'_e$ \cite{FN05}.

In that case, the symmetrized KL divergence (\ref{DSKLb}) simplifies to
\bea
D_S ( P , P') &=& \frac{1}{2} \sum_\alpha \left[ (\Delta S^+_\alpha)^2 - (\Delta S^-_\alpha)^2 \right] \nonumber\\
&=& \frac{1}{2}  \sum_{\alpha} (A_\alpha-A_\alpha') (J_\alpha-J_\alpha')  \, .
\label{DS.e} 
\eea
The separation between two models only depends on their thermodynamic currents and affinities. 
Here also the coordinates $S^{\pm}_e$ still vary along the joint light cone. 


The isoactivity and isokinetic manifolds have been studied in the context of nonequilibrium transport and response theory, where they are referred to as $e$ and $m$ equivalence classes, respectively. 
Along these manifolds, transport displays unique properties, including fully symmetric responses \cite{A12c, A22, V22, A24, A25}. In addition, the isokinetic manifold allows to change the output (set of currents) of a system while minimizing the disruption with respect to the original dynamics \cite{A24, A25}.
Remarkably, these two manifolds emerge naturally in this framework as the light cone of the symmetric variables.

\subsection{Entropy production as spacetime divergence}

The dissipation or entropy production 
\bea
\Delta_i S = \sum_\alpha J_\alpha A_\alpha \geq 0
\eea
plays an important role in thermodynamics \cite{NP77}. 
It is traditionally associated with the breaking of time-reversal symmetry \cite{G04}. 
Indeed, it can be expressed as the divergence with respect to the time-reversed chain $P^*$:
\bea
D_S \parent{P , P^* } = 2\, \Delta_i S 
\eea
since $P^*$ has the same activities and kinetic forces as $P$.
The time-reversed chain $P^*$ thus sits at the intersection of the two branches of the $e$ light cone. 

Interestingly, expression (\ref{DSKLb}) reveals that the entropy production also emerges as the divergence from other dynamics. 
Specifically, dynamics $P'$ satisfying $A'_\alpha =J'_\alpha =0$ together with either $Y'_e =Y_e$ or $X'_e =X_e$ would also lead to the entropy production. 
These conditions are satisfied at the intersection of the $e$ light cone and the equilibrium manifold, leading to two different solutions.

First, the equilibrium dynamics $P^e$ satisfies $X'_e = X_e$ so that
\bea
D_S \parent{P , P^e } = (1/2)\, \Delta_i S \, .
\eea
Second, the equilibrium dynamics $P_m = (P+P^*)/2$ satisfies $Y'_e = Y_e$ so that 
\bea
D_S \parent{P , P^m } = (1/2)\, \Delta_i S \, .
\eea
These last two formulas show that the entropy production can be understood as measuring the divergence from specific equilibrium dynamics \cite{A24b}.
In this sense, formula (\ref{DSKLb}) generalizes the previous results from Ref. \cite{A24b} and the second law inequality $\Delta_i S \geq 0$.
These expressions emerge from the underlying spacetime structure. 
They expand our understanding of irreversibility and have practical implications, such as new lower bounds \cite{A24b}.

\section{A molecular motor example} \label{MM}

Consider a Markov chain representing a molecular motor with $2\ell$ states corresponding to different conformations of the protein complex. 
These states form a cycle of periodicity $2\ell$ corresponding to a revolution by 360° for a rotary motor or a reinitialization step for a linear motor.
The motor alternates between two types of states according to the transition matrix \cite{AG06}
\bea
P=
\begin{pmatrix}
0 &  p_1 &  &  &  & 1-p_1\\
1-p_2 & 0 & p_2 &  &  & \\
 &  1-p_1 & 0 & p_1 & & \\
 &   & \ddots & \ddots & \ddots & \\
 &   &  & 1-p_1 & 0 & p_1\\
p_2 &   &  &  & 1-p_2 & 0
\end{pmatrix}_{2\ell \times 2\ell}  .
\nonumber
\eea
The matrix $P$ is doubly stochastic, so that its stationary state ${\bf \pi}= (1,1,\cdots ,1)/2\ell$ is uniform for all parameters $(p_1,p_2)$.
The average current $J$ and affinity $A$ thus take the form
\bea
J = \frac{1}{2\ell} \parent{p_1+p_2 -1}
\label{MM.J}
\eea
and
\bea
A = \ell \, \ln \frac{p_1p_2}{(1-p_1)(1-p_2)} \, .
\label{MM.A}
\eea
The system is at equilibrium when $J=A=0$ (i.e., when $p_2 = 1-p_1$). 

The activities read $Y_1 = (p_1 -p_2 +1)/2\ell$ and $Y_2 = (p_2 -p_1 +1)/2\ell$ depending if the transition is odd or even, respectively. 
Here, only one edge is independent due to the symmetry of the model and the constraint $Y_1+Y_2 =1/\ell$.
Taking 
\bea
Y = Y_1 = \frac{1}{2\ell}(p_1 -p_2 +1)
\label{MM.Y}
\eea
as the independent activity leads to an effective kinetic force
\bea
X = \ell \, (\bar{X}_1 - \bar{X}_2)  = \ell \, \ln \frac{p_1 (1-p_2)}{p_2 (1-p_1)} \, .
\label{MM.X}
\eea

The symmetrized KL divergence between two models $P = (p_1,p_2)$ and $P' = (p'_1,p'_2)$ reads
\bea
2 \, D_S (P,P')&=& (p_1-p'_1)\ln \frac{p_1}{p'_1} + (p_2-p'_2)\ln \frac{p_2}{p'_2} \nonumber \\
&+& (p'_1-p_1)\ln \frac{1-p_1}{1-p'_1} + (p'_2-p_2)\ln \frac{1-p_2}{1-p'_2} \, .
\nonumber
\eea
Using expressions (\ref{MM.J})-(\ref{MM.X}), a direct calculation leads to
\bea
D_S (P,P') = \frac{1}{2} (A-A') (J-J') + \frac{1}{2} (X - X') (Y - Y') \, .
\nonumber 
\eea
As expected from our general result (\ref{DSKLb}), the divergence is expressed in terms of the system's current and affinity on the one hand, and the activity and kinetic force on the other hand.

\begin{figure}[t]
\includegraphics[scale=.41]{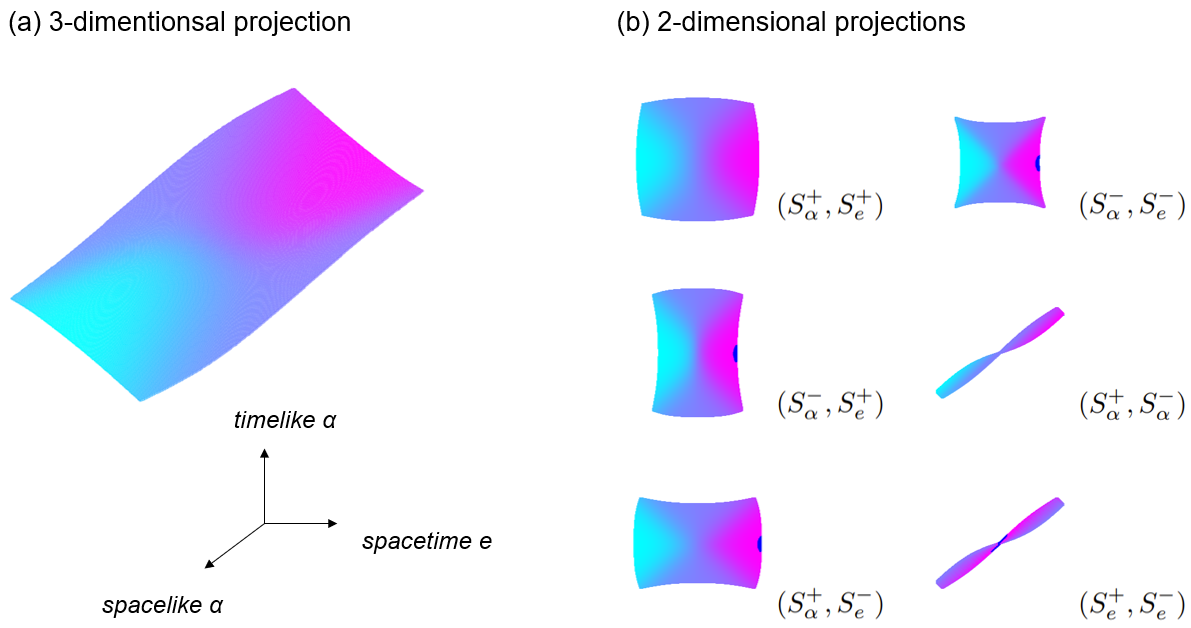}
\caption{{\bf Model manifold in Minkowski space for the molecular motor}.
The $2$-dimensional molecular model manifold is embedded into $2+2$ dimensions. 
(a) A three-dimensional projection of the manifold, which is colored based on the current $J$.
(b) The different two-dimensional projections of the manifold. 
The spacelike coordinates are denoted as $S^+$ and the timelike by $S^-$.
For visual clarity, the embedding coordinates are scaled as $S_\alpha^{\pm} = (A/\lambda \pm \lambda J)/2$ and $S_e^{\pm} = (X/\lambda \pm \lambda Y)/2$ with $\lambda = \ell$. All 2-d projections have the same scale. }
\label{fig2}
\end{figure}

The four embedding coordinates take the form
\bea
S_\alpha^{\pm}  = \frac{1}{2}(A\pm J) \, , \quad \quad  S_e^{\pm}  &=& \frac{1}{2}(X \pm Y) \, .
\eea
In these coordinates, the molecular motor models form a 2-dimensional manifold in a $2+2$ Minkowski space (Figure~\ref{fig2}).  
Note that the number of embedding coordinates is independent of the size $\ell$ of the motor.

At equilibrium where $p_1 = 1-p_2$, the current and affinity vanish so that $S_\alpha^{\pm} = 0$. The equilibrium manifold is thus entirely described by the coordinates $S_e^{\pm}$. 
When $p_1 = p_2 = p$, the motor is spatially homogeneous and the kinetic force $X = 0$ while the activity $Y = 1/2\ell$ is independent of $p$. 
The manifold $X_e = 0$ is thus described entirely by the coordinates $S_\alpha^{\pm}$. 

\begin{figure}[t]
\includegraphics[scale=.18]{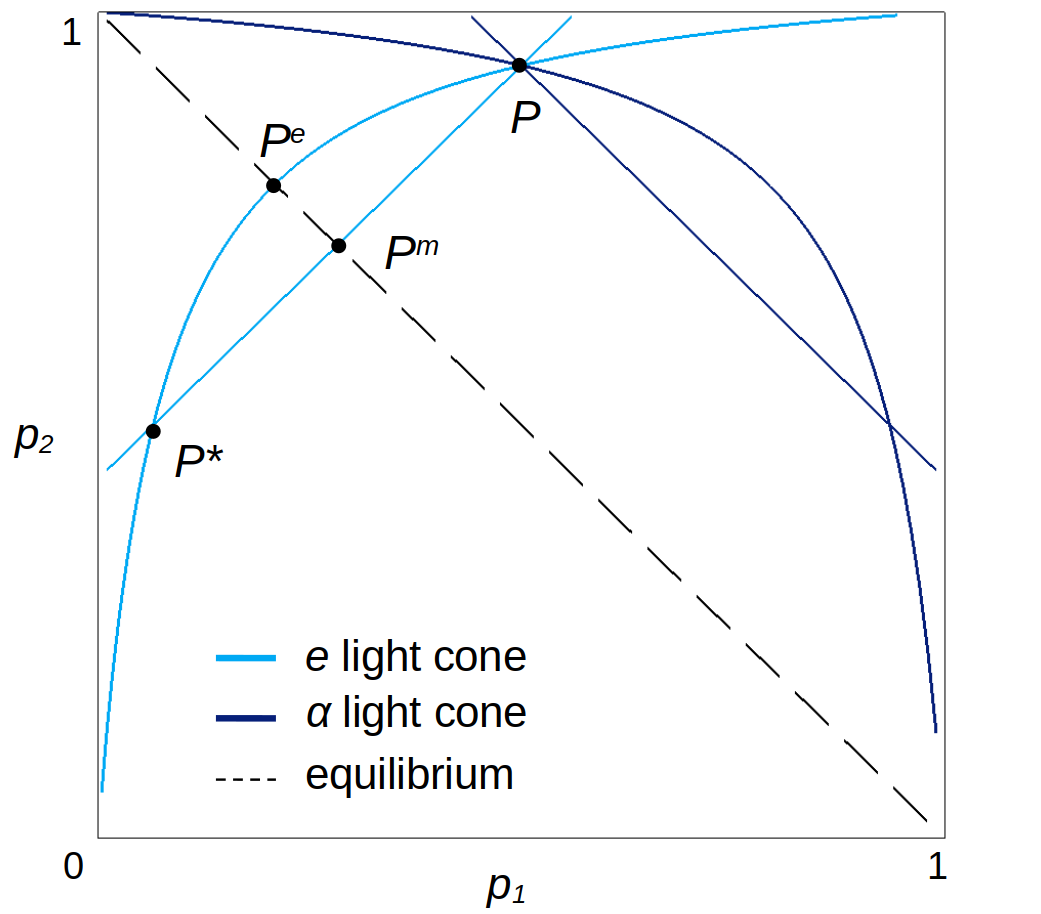}
\caption{{\bf Light cones in parameter space for the molecular motor}.
The $\alpha$ light cone is composed of the isocurrent and the isoaffinity branches, and the $e$ light cone 	is composed of the isoactivity and the isokinetic branches.
The entropy production is given by three possible divergences with respect to $P^*$, $P^e$ and $P^m$.
All three dynamics belong to the light cone $e$, with $P^*$ sitting at the intersection of the two branches while the latter two intersecting the equilibrium manifold (dashed line).
Expressions for $P^*, P^e$, $P^m$ and the light cones are provided in the main text.
}
\label{fig1}
\end{figure}

The $\alpha$ and $e$ light cones are each composed of two branches forming a straight and a curve path in parameter space (Figure~\ref{fig1}).
The obtain the isoaffinity and isokinetic manifolds, we express the model parameters as 
\bea
p_{1} = \frac{{\rm e}^{(A+X)/2\ell}}{1+{\rm e}^{(A+X)/2\ell}} \, , \quad \quad p_{2} = \frac{{\rm e}^{(A-X)/2\ell}}{1+{\rm e}^{(A-X)/2\ell}} \, . 
\label{MM.param} 
\eea
The isoaffinity manifolds are obtained by fixing $A$ and varying $X$ while the isokinetic manifolds are obtained by varying the affinity $A$ at fixed value of $X$.  
In turn, from expressions (\ref{MM.J}) and (\ref{MM.Y}) we see that the isocurrent and isoactivity manifolds form straight lines in parameter space.


On the $e$ light cone the transport properties display unique characteristics, including a well defined response curves, symmetric responses, and a direct link between equilibrium and nonequilibrium fluctuations. Exact expressions for transport along these manifolds can be found in reference \cite{A25}	.

The entropy production is obtained as the divergence from $P^* = P^T$, as well as from the equilibrium dynamics $P^{e}$ and $P^{m}$ so that \cite{A24b}
\bea
\Delta_i S= J \times A  = 2 \, D_S \parent{P , P^{e}}  = 2 \,  D_S \parent{P , P^{m}}\, .
\eea
Here $P^e$ is given by $p^e_1 = \sqrt{p_1(1-p_2)}/\Sigma, p^e_2 = \sqrt{p_2(1-p_1)}/\Sigma$ with $\Sigma = \sqrt{p_1(1-p_2)}+\sqrt{p_2(1-p_1)}$, and $P^{m}$ is given by $p^m_1 = (p_1-p_2+1)/2, p^m_2 = (p_2-p_1+1)/2$. 
$P^e$ and $P^m$ belong to the intersection of the $e$ light cone and the equilibrium manifold (Figure~\ref{fig1}). 

\section{Summary and outlook}

This work brings together thermodynamic and information geometric concepts, showing that thermodynamic quantities emerge naturally from the Minkoswki structure induced by the symmetrized KL divergence.
While other divergences could be used, the symmetrized KL divergence bypasses the curse of dimensionality and provides the lowest embedding dimension, at least for exponential families of probability distributions \cite{TQS20, QEA23}. 


The resulting light cones play a crucial role in nonequilibrium transport.
In particular, the light cones of the symmetric variables correspond to isokinetic and isoactivity manifolds. 
Along these two manifolds, nonequilibrium transport satisfy special properties, such as a fully symmetric response, relationships between average currents and their large fluctuations, and an equivalence between equilibrium and nonequilibrium fluctuations \cite{A12c, A22, V22, A24, A25}. 


The entropy production also emerges naturally as the spacetime divergence with respect to the time-reversed dynamics, but also to specific equilibrium dynamics at the intersection of the $e$ light cone and the equilibrium manifold \cite{A24b}.
The positivity of the entropy production is rooted in the positivity of the divergence; whether the more general inequality (\ref{DSKLb}) generates additional constraints for nonequilibrium phenomena remains an open question \cite{FN06}. In addition, concepts such as angles or the global light cones in Minkowski space may offer further physical insights \cite{FN04}.

This framework provides a systematic way to analyze the space of Markov models and to gain insights into the behavior and design of mesocopic devices.
The embedding can be used to determine the minimum parameters needed to describe model data or to identify the governing parameters for model prediction and coarse graining \cite{MEA13, QEA23}.
The spacetime structure could also help untangle important phenomena such as dynamical or nonequilibrium phase transitions. 

\vskip 0.5 cm


\section*{Appendix: Derivation of the model manifold and its Minkowski embedding}

The first step to obtain the Minkowski embedding is to express the symmetric divergence (\ref{DSKLb}) between two models in terms of their kinetic and thermodynamic variables. 
To this end, we first calculate the log ratios $\ln P_{ij}/P'_{ij}$ that appear in Eq. (\ref{DsKL}). 
Using the thermodynamic decomposition (\ref{P.decomp}) with $P = s[P^e\circ Z]$ and $P' = s[P^{'e}\circ Z']$, the log ratios take the form
\bea
\ln \frac{P_{ij}}{P'_{ij}} = \ln \frac{\rho'}{\rho} + \ln \parent{ \frac{\alpha_j'}{\alpha_i'} \frac{\alpha_i}{\alpha_j}}  + \ln \frac{Z_{ij}}{Z'_{ij}} + \ln \frac{P^e_{ij}}{P^{'e}_{ij}} \, .
\label{P/G}
\eea
The first term $\ln \rho'/\rho$ vanishes when inserted into expression (\ref{DsKL}) since $\sum \pi_i P_{ij} = \sum \pi'_i P'_{ij} = 1$. The second term vanishes as well when averaged over a stochastic dynamics (Lemma 4.3 (iii) in ref. \cite{WW21}). 
The third term in Eq. (\ref{P/G}) reads
\bea
(1/2) \sum_{\alpha} (A_\alpha-A_\alpha') (J_\alpha-J_\alpha')  	
\label{app.AJ}
\eea
when inserted into (\ref{DsKL}), where we used that $\ln Z_{ij} = \pm A_\alpha/2$ if the transition $i\rightarrow j$ corresponds to $\alpha$ in the positive (negative) direction, and $0$ otherwise. 

To calculate the last term $\ln P^e_{ij}/P^{'e}_{ij}$, remember that $P^e_{ij} = (1/\lambda)(\gamma_i/\gamma_j)\sqrt{P_{ij}P_{ji}\pi_j\pi_i}$ where $\lambda$ and $\gamma$ are the Perron eigenvalue and eigenvector of $\sqrt{P_{ij}P_{ji}\pi_j/\pi_i}$, and similarly for $P^{'e}$. 
Here also, the terms $\ln \lambda'/\lambda$, $\ln (\gamma'_j \gamma_i/\gamma'_i \gamma_j)$ and $(1/2)\ln (\pi'_j \pi_i/\pi'_i \pi_j)$ vanish when averaged over $\sum [\pi_i P_{ij} - \pi'_i P'_{ij}]$.
The remaining factor then becomes
\bea
(1/2) \sum_e  (\bar{X}_e - \bar{X}'_e) \parent{Y_e - Y'_e} \, ,
\label{app.XY}
\eea
where the sum runs over the (undirected) transitions $e = i \longleftrightarrow j$, $Y_e = \pi_i P_{ij} + \pi_j P_{ji}$ is the activity, and $\bar{X}_e = \ln P_{ij}P_{ji}$.

Note, however, that not all variables $Y_e$ are independent since $\sum_e Y_e = 1$.
One variable $r$ can thus be eliminated as $Y_r = 1 - \sum_{e \neq r} Y_e$.
Inserting this constraint into (\ref{app.XY}) and summing over the different terms, we have that
\bea
& & \sum_e  (\bar{X}_e - \bar{X}'_e) \parent{Y_e - Y'_e} \nonumber \\
&=& \sum_{e\neq r}  (\bar{X}_e - \bar{X}'_e) \parent{Y_e - Y'_e} + (X_r - X'_r) \sum_{e \neq r} (Y'_e - Y_e ) \nonumber  \\
&=& \sum_{e\neq r}  (\bar{X}_e-\bar{X}_r - \bar{X}'_e+\bar{X}'_r) \parent{Y_e - Y'_e} \nonumber \\
&=& \sum_{e\neq r}  (X_e - X'_e) \parent{Y_e - Y'_e} \, ,
\eea
where we introduced the effective forces 
\bea
X_e = \bar{X}_e - \bar{X}_r \, .
\eea 
This substitution leads to Eq. (\ref{DSKLb}).

Next, to obtain the embedding coordinates (\ref{icoord}), each term in expression (\ref{app.AJ}) can be written as \cite{TQS20}
\bea
(A_\alpha-A_\alpha') (J_\alpha-J_\alpha') = (\Delta S^+_\alpha)^2 - (\Delta S^-_\alpha)^2
\eea
with the two coordinates
\bea
S^{\pm}_\alpha = \frac{1}{2} (A_\alpha \pm J_\alpha) \, .
\eea
The same factorization applies to the terms $(X_e - X'_e) (Y_e - Y'_e)$ 
with the two coordinates
\bea
S^{\pm}_e =  \frac{1}{2} (X_e \pm Y_e) \, .
\eea

 



\begin{thebibliography}{99}
\bibitem{FN01} While $D_S (P,P')$ is symmetric, non-negative, and vanishes only when $P = P'$, it doesn't define a distance since the triangle inequality is not always respected.
\bibitem{FN04} The joint light cone is given by the condition $\sum_\alpha (A_\alpha-A'_\alpha)(J_\alpha-J'_\alpha) = 0$. However, this set doesn't appear to have a clear physical interpretation, so here we focus on the intersection of the individual $\alpha$ light cones.
\bibitem{FN05} When using the dependent kinetic forces $\bar{X}_e$, the joint light cone is described by the condition $\bar{X}_e - \bar{X}'_e = C$, with $C$ a constant. Indeed, we then have $\sum_e (\bar{X}_e - \bar{X}'_e) (Y_e -Y'_e) = C \sum_e (Y_e -Y'_e) = 0$ since $\sum_e Y_e = \sum_e Y'_e = 1$. The condition $\bar{X}_e - \bar{X}'_e = C$ was used in previous work to describe the $e$ equivalence class \cite{A12, A22, A25}.
\bibitem{FN06} For instance, the inequality $\sum_e \bar{X}_e Y_e \leq 0$ for the symmetric variables is reminiscent of the entropy production $\Delta_i S = \sum_\alpha J_\alpha A_\alpha \geq 0$ for the antisymmetric variables. Its potential role in understanding nonequilibrium phenomena remains to be investigated.
\end{thebibliography}
\end{document}